# Mapping for accessibility: A case study of ethics in data science for social good

Anissa Tanweer
University of Washington
Seattle, WA, USA
tanweer@uw.edu

Margaret Drouhard
University of Washington
Seattle, WA, USA
mdrouhar@uw.edu

Brittany Fiore-Gartland
University of Washington
Seattle, WA, USA
fioreb@uw.edu

Nicholas Bolten
University of Washington
Seattle, WA, USA
bolten@uw.edu

Jess Hamilton
University of Washington
Seattle, WA, USA
jesshami@uw.edu

Kaicheng Tan
University of Washington
Seattle, WA, USA
tan95@cs.washington.edu

Anat Caspi
University of Washington
Seattle, WA, USA
caspian@cs.washington.edu

## ABSTRACT

Ethics in the emerging world of data science are often discussed through cautionary tales about the dire consequences of missteps taken by high profile companies or organizations. We take a different approach by foregrounding the ways that ethics are implicated in the day-to-day work of data science, focusing on instances in which data scientists recognize, grapple with, and conscientiously respond to ethical challenges. This paper presents a case study of ethical dilemmas that arose in a "data science for social good" (DSSG) project focused on improving navigation for people with limited mobility. We describe how this particular DSSG team responded to those dilemmas, and how those responses gave rise to still more dilemmas. While the details of the case discussed here are unique, the ethical dilemmas they illuminate can commonly be found across many DSSG projects. These include: the risk of exacerbating disparities; the thorniness of algorithmic accountability; the evolving opportunities for mischief presented by new technologies; the subjective and value-laden interpretations at the heart of any data-intensive project; the potential for data to amplify or mute particular voices; the possibility of privacy violations; and the folly of technological solutionism. Based on our tracing of the team's responses to these dilemmas, we distill lessons for an ethical data science practice that can be more generally applied across DSSG projects. Specifically, this case experience highlights the importance of: 1) Setting the scene early on for ethical thinking 2) Recognizing ethical decision-making as an emergent phenomenon intertwined with the quotidian work of data science for social good 3) Approaching ethical thinking as a thoughtful and intentional balancing of priorities rather than a binary differentiation between right and wrong.

## 1.INTRODUCTION

Ethics in the emerging world of data science are often discussed through cautionary tales about the dire consequences of missteps taken by high profile companies or organizations. Instead, we focus on instances in which data scientists recognize, grapple with, and conscientiously respond to ethical challenges, foregrounding the ways that ethics are implicated in the quotidian work of data science. This paper grew out of collaborations between a team of ethnographers studying the phenomenon of data science for social good and a team of participants in a Data Science for Social Good (DSSG) program at the University of Washington. After describing the program's approach to ethics, we provide an account of the ethical dilemmas that arose in the course of executing one of the projects supported by that program, one that is focused on improving navigation for people with limited mobility. We describe how the project team responded to those dilemmas, and the ways in which those responses gave rise to still more dilemmas. We contend that the ethical dilemmas the team faced are not rare, but common across a variety of data science for social good applications. Therefore, we articulate how these dilemmas were addressed in a particular DSSG project in order to draw more general lessons that can be applied toward the development of ethical practices in data science for social good efforts. These lessons highlight the importance of: 1) Setting the scene early on for ethical thinking 2) Recognizing ethical decision-making as an emergent phenomenon intertwined with the quotidian work of data science for social good 3) Approaching ethical thinking as a thoughtful and intentional balancing of priorities.

Each author on this paper has been involved with the Data Science for Social Good (DSSG) program run by the eScience Institute at the University of Washington, which has been described in a previous Bloomberg D4GX paper [16]. The ethnographers (Tanweer, Drouhard, Fiore-Gartland) have collectively spent approximately 1,400 hours observing data science for social good projects in action, and have conducted more than 100 interviews with DSSG participants. As participant-observers, they are also directly involved in the development of UW's DSSG program. This includes, among other things, conducting workshops with participants about ethics, stakeholders, and collaboration that helped shape the trajectory of the project we discuss herein. The authors of this paper who are DSSG practitioners (Drouhard, Bolten, Hamilton, Tan, Caspi) have spent the last three years developing a routing application for people with mobility impairments called AccessMap [4]. In the course of this work, they also formed a complementary effort called OpenSidewalks to develop data standards, tools, and practices for creating an open-source pedestrian map layer that would serve as the foundation for AccessMap. Given the tight coupling of the AccessMap and OpenSidewalks objectives and team membership, the group will hereafter be referred to as the AccessMap/OpenSidewalks (AMOS) team. One of the ethnographers (Drouhard) spent a summer embedded as a member of the AMOS team, fully integrated into their day-to-day work and directly contributing to the project's goals as a designer and programmer.

The AMOS team aimed to build routing software that would meet the specific needs of individuals with limited mobility. The application was to account for aspects of the built environment that influence travel decisions for people who use wheelchairs or have other limitations in their ability to navigate the built environment. This includes information such as the location of sidewalk curb cuts, the presence of crosswalks, the steepness an incline, and the surface quality of a pedestrian path. Achieving this goal involved a number of stages typical in the development of many DSSG projects. The team first had to **develop a foundational infrastructure**, which in their case meant building a pedestrian map layer, or a connected graph of features such as sidewalks and street crossings that a routing algorithm could traverse. The team also had to engage in **data design**, which often entails figuring out which data to include and how to best represent it. For the AMOS team, this meant deciding which information would be most salient to people with mobility impairments and which standards should be developed for depicting this information. Another stage of development is **deployment and adoption**, in which the AMOS team had to consider how to build customizable profiles and user authentication. As we will describe below, in each stage of development, the solutions they devised to meet the project's needs each came with its own set of ethical considerations that demanded the team's attention, affected the design of their project, and ultimately gave rise to new challenges or dilemmas with yet another set of ethical considerations.

While the details of the AMOS team's DSSG project are unique, the ethical dilemmas they illuminate can be routinely found across many data science for social good projects. These issues include: the risk of exacerbating disparities; the thorniness of algorithmic accountability; the evolving opportunities for mischief presented by new technologies; the subjective and value-laden interpretation at the heart of any data-intensive project; the potential for data to amplify or mute particular voices; the possibility of privacy violations; and the folly of technological solutionism.

## 2. CASE STUDY

### 2.1. INFRASTRUCTURE DEVELOPMENT: THE AMOS TEAM DEALS WITH ISSUES OF DISPARITY, ALGORITHMIC ACCOUNTABILITY, AND MISCHIEF

The AMOS team felt that OpenStreetMap (OSM) could provide the ideal platform for creating the requisite pedestrian layer they would need to build AccessMap. As an open source project, OSM was free to use and potentially flexible enough to work for their purposes. Although OSM is available worldwide, and the community aspires to one day have the entire globe thoroughly mapped, its crowd-sourced data is currently unevenly distributed. Some geographic areas are covered with detail and precision while other areas have been barely mapped at all; likewise, some features of the built environment (such as roads) tend to be more thoroughly documented than others (such as buildings).

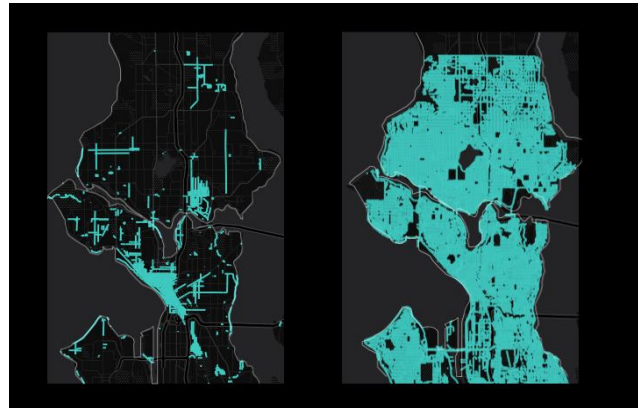

Fig. 1. Turquoise lines represent mapped sidewalk segments for the city of Seattle. The image on the left depicts sidewalk data coverage as it existed in OpenStreetMap when the AMOS team began their project. The image on the right depicts the sidewalk coverage that would exist if they imported municipal open data into OpenStreetMap.

When the team was considering whether or not they should use OSM to build AccessMap for the City of Seattle, information about Seattle's sidewalks in OSM was highly concentrated in the downtown area, and largely absent from the peripheral neighborhoods (Fig. 1). If the team relied solely on this crowdsourced data created by individual OSM contributors, they would not have exhaustive coverage of sidewalks. Knowing that the urban core already had more amenities and informational resources available for people with limited mobility compared to most of the city, they quickly realized that this could serve to further entrench or exacerbate existing inequities. This particular ethical dilemma they faced is a common one in DSSG projects, in which the reliance on conveniently available data can serve to **create or compound disparities** [3,5].

The solution they proposed for creating a more complete sidewalk graph, and therefore more equitable information resource, was to import the City of Seattle's publicly available sidewalk data. By including that data, their sidewalk graph would cover the entire city fairly comprehensively (Fig 1). However, this database was maintained for inventory purposes rather than mapping purposes. Therefore, the coordinates provided for each individual sidewalk segment were only approximations, and in most cases, the sidewalk segments didn't line up perfectly. Instead, there were gaps between segments that precluded creating a connected graph for a routing algorithm to traverse. So the team wrote algorithms to clean the data and align the segments (Fig. 2). This resulted in a connected sidewalk layer that could be added to OSM as the foundation for their routing algorithm.

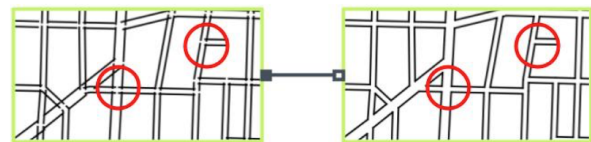

Fig. 2. The image on the left depicts imprecise geolocation data for sidewalk segments in Seattle's open municipal dataset. The image on the right depicts sidewalk segments in a connected graph after the AMOS team algorithmically cleaned the municipal dataset.

But while this solution was now technically feasible, they faced another dilemma: the OSM community strongly discouraged mass imports of data like the one the team was hoping to do. According to prevailing norms in OSM, the map should be built and curated by a community of human contributors based on their physical observation of the built environment. In part, this is a philosophical ideal, but is also informed by technical limitations. OSM stores every version of data that is ever entered into the map. If incorrect data is entered, and later reverted or corrected, both the incorrect and correct version will forever be stored on OSM's servers. One concern with programmatically cleaned and imported data, then, was that if the data were wrong and needed to be changed, the organization would end up storing massive amounts of useless data and straining their computational resources. For this reason, OSM standard practice was that human contributors independently verify every piece of information in the map.

These concerns raised yet another ethical issue commonly encountered in DSSG efforts, that of **algorithmic accountability**. Algorithmic accountability is often discussed in the context of algorithm-assisted decision-making, and the potential for algorithms to result in biased or discriminatory outcomes [3,20]. For example, recent journalism and scholarship has discussed how predictive machine learning algorithms used in the criminal justice system can unfairly affect people of color [2,17].

As such, much of the conversation around algorithmic accountability centers on justifiable and much-needed calls to improve the process of algorithm development and deployment, including correcting for biased training data, and make the design of publicly-relevant algorithms transparent [1,6,15]. But the AMOS team's experience with OSM points to a broader view of algorithmic accountability -- one that means acknowledging how algorithms and programmatic tasks are enmeshed in complex politics, histories, and cultures [8,11]. It means understanding their role in complex socio-technical systems where human agency and computational agency are tightly intertwined. It means working with community norms and values, and compromising on when it makes sense to privilege human judgment and when it is appropriate to rely on computational automation.

This is exactly what the AMOS team did. In navigating the challenge of accountability and verification, they adopted an approach that had been used in the past by other groups with similar goals of importing mass data in humanitarian crisis zones or from municipal data sets. They adapted a tasking manager to import data in smaller chunks (rather than doing a single mass import), and created tasks for online users to view and validate these smaller changesets. They organized mapathon events to concentrate validation efforts and provide training in using the tasking manager. They also tried to fulfill their value for algorithmic accountability by making their data cleaning and import tools open source and available for others to review.

However, their respect for the OSM community norms that valued human judgment and contribution posed yet another dilemma that affected their relationship with another group of stakeholders in their project: government agencies. One of the AMOS team's long-term goals is to get municipalities to think of OSM as a shared data commons that they are invested in curating because they could use the data for planning and operating purposes. But the team also recognized that governments have serious concerns about data quality and integrity, and so they're hesitant to trust crowdsourced data because people can easily provide false information. And this does sometimes happen in OSM. For example, some OSM contributors have noticed a recent increase in the mislabeling of certain features in the built environment. They're finding features labeled on the map as public ponds that are actually private swimming pools, and public parks that are actually someone's front yard. They suspect, but don't know for sure, that Pokémon Go is using OSM for the location of certain landmarks, and so some players are gaming the system by falsifying features that would allow them to set up a lure at their own house and accrue more points. This is exactly the kind of behavior that would make government agencies hesitant to invest in, contribute to, or rely on OSM. And it represents another common ethical dilemma in DSSG efforts; the phenomenon that nearly every technological innovation is accompanied by new **opportunities for mischief or malice**. For this reason, it is important for developers and designers in the DSSG space to realize that down the road, the technologies they design may be used by different actors in different contexts and in different ways than intended [10]. And while they may not be able to foresee *all* of those possibilities, it is imperative to consider the design of technological affordances in order to mitigate harm or avoid ethical missteps in the future. With this in mind, the AMOS team has thought deeply about how to deal with issues of data quality. Their solution is to advocate for a "data commons" infrastructure that allows organizations to retain a protected copy of their data and imposes data validation processes to ensure data integrity. They detail their ideas for this design in a forthcoming article appearing in the IBM Journal of Research and Development [4].

## 2.2. DATA DESIGN: THE AMOS TEAM DEALS WITH ISSUES OF INTERPRETATION AND VOICE

Aside from establishing a base layer for their application, the AMOS team needed to populate that layer with features and annotations that are relevant to people with limited mobility, such as the presence of curb cuts, surface composition and quality, and the steepness of inclines. However, OSM mapping standards had been developed primarily with automobile travel in mind, rather than pedestrian travel, and so sidewalks were documented only as attributes of streets, rather than as features on the map in their own right. Essentially, they were considered to be metadata, and their topographies and features were incomplete. That made it hard to map sidewalks with precision, and even harder to annotate sidewalks with the kinds of information the OpenSidewalks team wanted to include on the map. For their own purposes, and for the purposes of future pedestrian-centric mapping applications, the team saw the need for sidewalks to be included not just as attributes of streets, but as first-class features of the map that could be precisely located and intuitively annotated with information relevant to pedestrian travel. So their solution was to try changing the OSM standards.

But that, of course, also presented a dilemma. The AMOS team knew that they were operating from a very particular position, privileging a particular kind of information for a particular purpose, and that there was sure to be some resistance from at least some segments of the OSM community, which is distributed around the world and operates democratically. In this case, some people in the OSM community were hesitant to separate sidewalks from named streets because they thought it could be difficult to produce verbal, turn-by-turn routing instructions without the street names. The team was now confronted with yet another common dilemma in DSSG efforts: their **subjective, value-based interpretation** of which data mattered and why it mattered was not universally shared [7,9].

Their solution began with recognizing that they couldn't simply impose their own interpretation and values onto this community. Instead, they took the time to learn about the communities' own narratives and interpretations of why it existed and what was important. They spent weeks combing through hundreds of pages of discussion threads and listserv archives to understand how the OSM community made decisions and what it valued. They sought out leaders in the local OSM community to collaborate on local mapping projects and to get advice on how best to proceed. The better they understood the OSM community and sub-communities, the more they were able to incorporate the community's conceptualizations of value into their own DSSG project. And the more time they spent engaging with the community on its own terms and through its own channels of communication, the more trust and credibility the team was able to build. If they hadn't taken the time to understand the values of the community and subsequently refine their own approach, their interpretation of the problems with the current data collection standards and their proposed solutions likely would have seemed an alien imposition to the OSM community. However, when they finally presented their proposal to audiences at a national OSM conference and in online discussion spaces, their ideas were largely well-accepted and supported.

Even with that success, the AMOS team still had to decide what kinds of information to include and exclude in AccessMap, and how to present that information. Very early in the process of developing the idea for AccessMap, the team interviewed a number of people with limited mobility to find out what their informational needs were. This led to ongoing relationships with a several of these individuals, who became key informants and collaborators throughout the development process. One manual wheelchair user they worked with considered information about curb cuts to be fairly unimportant, since he was able to roll over curbs several inches high. For him, steepness of the sidewalk itself was the most important factor that affected his route decisions, because he had to use his own energy and upper body strength to get up hills. But they also talked with some powered wheelchair users who had the exact opposite view: they could make it up and down just about any hill, but their chairs could easily tip over if they attempted to navigate a curb that was more than a few centimeters high. The team also realized that while they had been considering the needs of people with mobility limitations, they hadn't considered the distinct needs of people with impaired vision. This process of investigation highlighted how decisions in data design would validate and privilege certain experiences and perspectives, while omitting others [18].

The AMOS team had long discussions and debates about different data structures and representations that privileged particular use cases. Ultimately they realized that there was no overwhelmingly superior design decision that could accommodate all types of users and use cases; each decision would inevitably prioritize some values over others. However, they still found that thinking through these decisions and consciously deciding which values to prioritize in particular cases produced a more flexible, equitable, and overall better design than basing decisions on ease of implementation or the convenience of pre-existing data structures. And in their case, the team did their best to accommodate diverse users by enabling customization of user profiles so that individuals could select for the features that were most important to them.

## 2.3. Deployment and adoption: The AMOS team deals with issues of privacy and security

Similar to every other solution they devised, this one presented yet another dilemma with ethical implications. The creation of user profiles would allow people of different abilities to customize their preferences and constraints, and therefore better serve the needs of their target audience. But this also introduced a number of concerns surrounding **privacy and security** because the team would have to make sure they were authenticating legitimate users and safeguarding their login credentials. Creating a system of authentication from scratch would be cumbersome, time-consuming, and unfeasible for a project that is supported largely by student labor. With robust authentication protocols already existing in the commercial space, the team's leadership had to decide whether it made sense to have students work on recreating a solution to a problem that has already been solved, at the expense of not spending their time working on novel applications. Ultimately, they made the pragmatic decision to use the widely-adopted open authorization standard OAuth, which would make it possible for users to log on to their application using their credentials for other services such as Facebook and Google. As such, they had to think about their AccessMap application not as a stand-alone product, but as a tool enmeshed in a larger socio-technical ecosystem with imperatives for interoperability. Although this renders their app more tractable from a development perspective, and more convenient from a user perspective, it also means that their work inevitably inherits the ethical perspectives and practices implicated in that wider ecosystem. To some extent, then, they recognized that there are limits on their abilities to exert their own ethical judgments on this system. What they can do is find a way to clearly, meaningfully, and appropriately communicate to their users exactly what data is being exchanged across this ecosystem, and allow their users the greatest control possible over their data and preferences. Currently, this is one of the major areas under development.

## 2.4. The coevolution of problem and solution: Dealing with the issue of technological solutionism

Even as the AMOS team deliberated and addressed each of these dilemmas in turn during the course of working on their project, they realized that there was one overarching dilemma with important ethical implications. There is a danger in technologists ignoring the complex, deep-rooted causes of social problems in order to propose simple technological fixes, what Morozov has called the "**folly of technological solutionism**" [12]. Aware of this trap, the AMOS team imagined AccessMap not as an intervention that could "solve" mobility for people with disabilities, but more modestly as an attempt to fill a known informational gap for a specific subset of the population that has been historically marginalized. They understood that this marginalization ran deep, and that their app couldn't resolve it. As students and volunteers came on board for short-term stints on the project over the course of several years, the project's leaders made a point of educating this rotating cast on the systemic roots of mobility impairment, and tried to help them understand disability as a social construct. In other words, the team adopted the perspective that disablement is not an inherent quality of individuals, but rather arises from a built environment that isn't designed with diverse needs in mind; if designs accommodated an exhaustive range of needs, people of different abilities would be

"able" to access and navigate the built environment with greater ease [13,14,19]. For example, if a door is designed to open with a lever handle rather than a round knob, people who can't grip well due to arthritis or some other limitation can still open the door on their own; they are not rendered disabled in that situation.

Given this perspective, while the AccessMap application was designed to help users navigate the built environment as it stood, the team also looked for ways to affect how people think about the relationship between disability and the built environment, and to advocate for policies that substantively increase accessibility. For example, the team used some of the same data that informed their app to show how a change to elevator access in downtown Seattle would affect travel options for people with limited mobility. Downtown Seattle is built atop steep hills, with the incline of some blocks approaching 20 percent (by comparison, the Americans with Disabilities Act mandates that wheelchair ramps be no steeper than 4.8 degrees). Because it's impossible for many individuals with mobility limitations to traverse such steep blocks, people have figured out how to use the elevators in buildings to go from one level on one side of a building to another level on the other side of that building, bypassing particularly steep blocks whenever possible. Unfortunately, however, not all buildings are available for this use, and some are only available during business hours. So the team created an isochrone visualization to demonstrate how access to elevators expands mobility options, and used that visual to advocate for building owners and managers to make their elevators publicly available at all times of day (Fig 3). This is just one of the ways that the team is purposefully leveraging their work to address the more systemic issues that perpetuate disability.

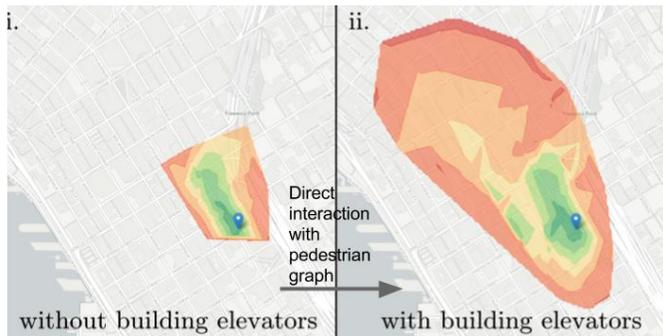

**Fig. 3.** These isochrone maps show accessible paths for an Access Map user with a manual wheelchair profile. The first map presumes buildings in the area do not make their elevators available for public use, while the second assumes that these elevators are publicly available. Each color indicates a particular level of difficulty for this user.

## 3. DISCUSSION

We have described above how the AccessMap/OpenSidewalk team encountered a number of ethical dilemmas throughout the various stages of their DSSG project, and how their efforts to resolve each of these gave rise to new dilemmas that demanded close consideration. It is our belief that the types of ethical challenges faced by the AMOS team are routine across many attempts to use data and data science in addressing social problems, and that the experience of the AMOS team can be instructive for approaching ethics in data science for social good applications more broadly. Below, we discuss how the AMOS experience highlights the importance of: 1) Setting the scene early on for ethical thinking 2) Recognizing ethical decision-making as an emergent phenomenon intertwined with the quotidian work of data science for social good 3) Approaching ethical thinking as a thoughtful and intentional balancing of priorities.

### 3.1. SETTING THE SCENE EARLY ON FOR ETHICAL THINKING

In the DSSG program generally, and the AccessMap/OpenSidewalks project specifically, much was done to create the time, space, and tone that would be conducive to recognizing and discussing ethical implications of the work, and weaving these conversations into the fabric of the project. This included initiating and facilitating conversations that were explicitly about ethics at the outset of the DSSG program, and also taking the time to sensitize new team members to the historical, political, and cultural context of the problem they were working on. First, organizers of the DSSG program worked with the ethnographers to incorporate workshops into the beginning of the program that addressed ethics and stakeholder relationships. In these workshops, DSSG participants explored ethical dilemmas across high-profile data science projects, identified stakeholders in their own projects, and had discussions in their teams about the benefits, risks, and ethical implications of their work.[a] The purpose of these workshops was not to provide an exhaustive treatment of these topics, but rather, to establish a common vocabulary, create a safe space for raising concerns, and establish a precedent for thinking about the broader implications of the work. These workshops initiated some of the first conversations the AMOS team had about the ways their priorities and values intersected with the priorities and values of other stakeholder groups, conversations that shaped the nature and direction of the project in important ways. Second, the AMOS team's leadership made an effort to sensitize new team members to issues of disability, accessibility, and inclusivity. With team members coming from a variety of disciplinary backgrounds and life experiences, they worked to foster a shared understanding of disability as social construct, accessibility as a moral imperative, and inclusivity as a value that deeply informed everything they did. This framing helped the AMOS team engage with their work not only as an interesting technical challenge, but also as a profoundly ethical and political statement.

### 3.2. RECOGNIZING ETHICAL DECISION-MAKING AS AN EMERGENT PHENOMENON

In this setting, with discussions of ethics being foregrounded at the outset of the program and the project's leaders encouraging the team to view the broader social implications of their work, ethics did not surface as discrete decision-points precipitated by a crisis or conflict. Rather, ethical decision-making was integrated into every phase of the AMOS team's work. In fact, most members of the group did not necessarily think of themselves as "practicing ethics"; instead they thought simply that they were doing what was necessary to do their work well. That meant spending considerable time deliberating about the ramifications and implications of various options, and expending considerable energy and resources to understand their stakeholder communities. The thinking, deliberation, and communication that went into this was not treated as a process orthogonal to the work of data science, but as an essential component of data science. So rather than rushing through the deliberations and pushing the team members (who were working on a short 10 week timeline) to produce a tangible outcome, the team's leadership valued the time spent in lengthy discussions among the team and with stakeholder groups. At times, however, some of the AMOS team members expressed frustration with spending so much time "just talking" and eagerness to get to the "work" of writing code. Importantly, then, although the team's leadership prioritized and foregrounded the communicative and deliberative labor that went into their ethical decision-making process, there is a balance to be struck between the value of this work and the value of producing code for immediate use. With the right balance, the significant time and energy required to support ethical decision-making can be readily recognized as an integral and crucial part of the work of data science for social good.

---

[a] Slides for a version of these presentations are available on the University of Washington Data Science Studies SlideShare site: https://www.slideshare.net/DataScienceStudieseS

### 3.3. APPROACHING ETHICAL THINKING AS A THOUGHTFUL AND INTENTIONAL BALANCING OF PRIORITIES

Just as the AMOS team members sometimes resented how long it took to engage in ethical deliberations, they were sometimes frustrated by "going back and forth" between various design alternatives to evaluate how each design privileged particular needs and sets of data over others. As with many design decisions, there is rarely one obviously superior ethical choice. Rather, the AMOS team found that they would have to prioritize some values over others and work within the ecosystem of existing infrastructures and communities. As one of the AMOS project leads pointed out, DSSG projects often inherit the ethical implications from earlier design choices. Perhaps more importantly, the AMOS team recognized that future adaptations of the AMOS work and perhaps even other DSSG projects may inherit the ethical implications of the choices they made. Not only would this shape data structures, system interactions, and user experiences, the AMOS team's design process is likely to impact the approach future designers take in addressing similar ethical issues. Recognizing the potential for their choices to shape future work, the AMOS team made a commitment to an open and transparent design process. While they recognized that prioritization of some values could mean sacrificing others, they worked to make those decisions consciously and thoughtfully, and they also tried to ensure that their choices would be open to inspection and suggestions from stakeholders.

### 4. CONCLUSION

This paper demonstrates how ethics are implicated in the quotidian practices of data science through tracing how one DSSG project team recognized, grappled with, and responded to ethical challenges as they continued to emerge throughout the stages of project development. The ways in which the AMOS team investigated, navigated, and ultimately responded to these ethical dilemmas throughout infrastructure development, data design, deployment and adoption, as well as with the co-evolution of problem and solution are potentially instructive for other DSSG projects that share project features and challenges. Drawing from this work we outline three recommendations for cultivating a culture that supports ethical thinking in DSSG projects. for doing DSSG. The following recommendations, 1) Setting the scene early on for ethical thinking 2) Recognizing ethical decision-making as an emergent phenomenon intertwined with the quotidian work of data science for social good 3) Approaching ethical thinking as a thoughtful and intentional balancing of priorities, can help support and integrate the kinds of conversations and reflexivity necessary for developing robust ethical thinking across DSSG projects more broadly.

### 5. ACKNOWLEDGMENTS
This work was supported in part by the Gordon and Betty Moore Foundation, the Alfred P. Sloan Foundation, and the University of Washington's eScience Institute. We would like to thank our research participants and the many individuals who have contributed to the development of the Access Map and Open Sidewalks projects.